\documentclass[pra,twocolumn]{revtex4}
\usepackage{graphicx}
\begin{document}

\bibliographystyle{unsrt}

\date{\today}

\title{Optically enhanced production of metastable xenon}
\author{G.T. Hickman}
\author{J.D. Franson}
\author{T.B. Pittman}
\email{todd.pittman@umbc.edu}
\affiliation{Physics Department, University of Maryland Baltimore
County, Baltimore, MD 21250}

\begin{abstract}

Metastable states of noble gas atoms are typically produced by electrical discharge techinques or ``all-optical'' excitation methods. Here we combine electrical discharges with optical pumping to demonstrate ``optically enhanced'' production of metastable xenon (Xe*). We experimentally measure large increases in Xe* density with relatively small optical control field powers.  This technique may have applications in systems where large metastable state densities are desirable.

\end{abstract}

\pacs{xxxx}

\maketitle

Metastable noble gas atoms are of interest in fields ranging from radioactive dating \cite{lu2014} to Bose-Einstein condensation \cite{vassen2012}. They have also been identified as promising alternatives to ground-state alkali atoms for realizing strong photon-atom interactions in waveguide and cavity-based systems \cite{pittman2013,hickman2015}. In nearly all of this work, the metastable states are produced using RF or DC discharge techniques \cite{chen2001}.  A radically different approach involves ``all-optical'' production of metastable states, in which a resonant vacuum ultraviolet (VUV) photon induces a transition from the ground state, followed by two resonant near infrared (NIR) photon transitions to the metastable state \cite{young2002}.  This technique has been successfully used to produce metatstable krypton (Kr*)  \cite{young2002,ding2007,kohler2014}, and can be adapted for other noble gas atoms as well. In this paper, we investigate a related technique for the production of metastable xenon (Xe*).

\begin{figure}[t]
\includegraphics[width=3.0in]{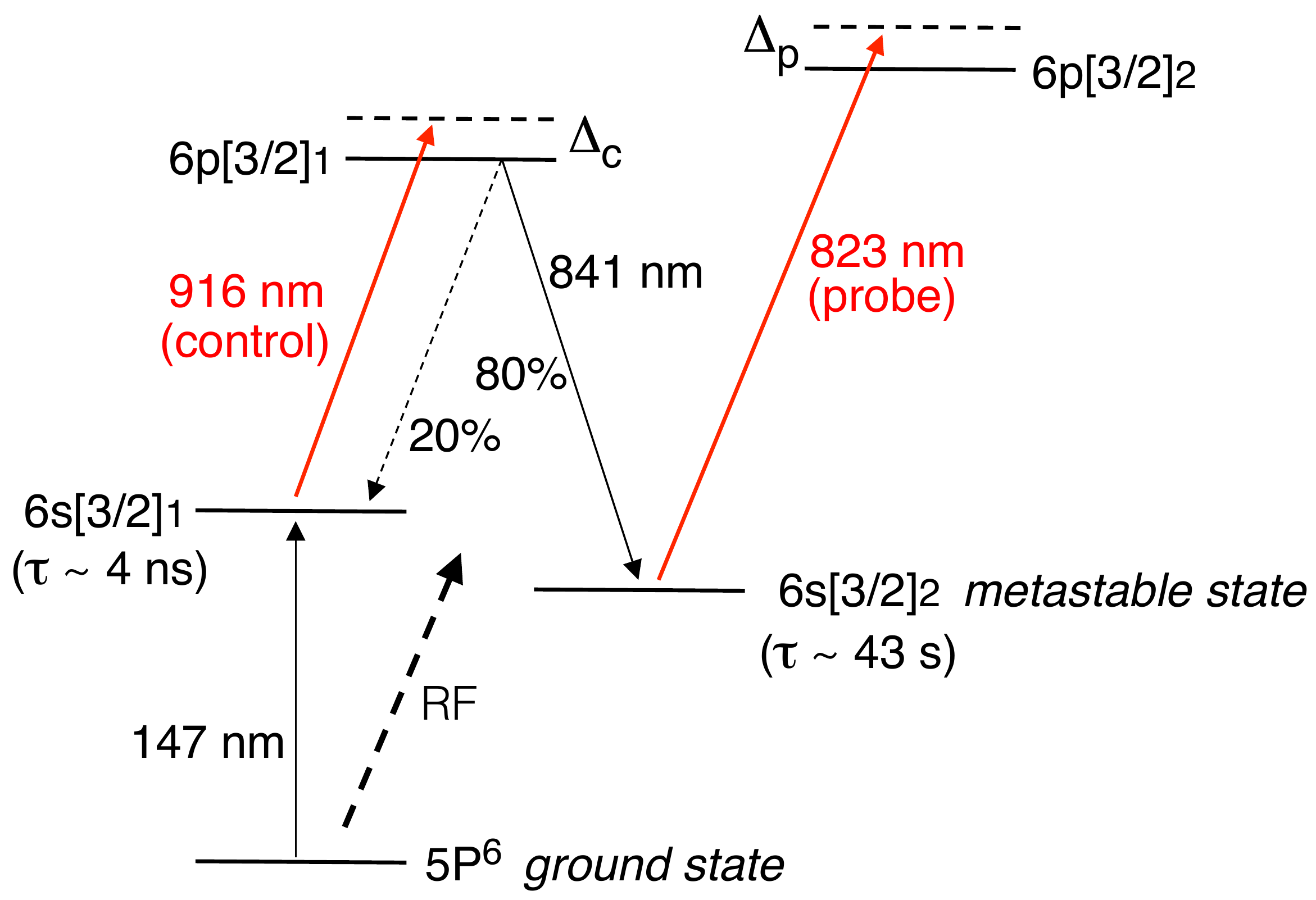}
\caption{(Color online) Overview of the relevant energy levels in xenon.  The $6s[3/2]_{2}$ metastable state density is dramatically enhanced by optically pumping population from the  $6s[3/2]_{1}$ auxilliary state.}
\label{fig:fig1}
\end{figure}

An overview of the xenon system is shown in Figure \ref{fig:fig1}. The atom can be optically excited from the ground state to the $6s[3/2]_{1}$ auxilliary state with a VUV photon at 147 nm. A control photon at 916 nm then causes a transition to the $6p[3/2]_{1}$ state, which subsequenty decays via spontaneous emission of an 841 nm photon to the $6s[3/2]_{2}$ metastable state with a branching ratio of $\sim$80\% \cite{NISTASD}. We then study the metastable state population by measuring absorption of an auxilliary 823 nm probe on the $6s[3/2]_{2} \rightarrow 6p[3/2]_{2}$ transition.

Due to the unavailability of VUV lasers, the generation of a high-flux of narrowband resonant 147 nm photons is problematic. In a true ``all-optical'' system \cite{young2002}, these photons would be produced by an external Xe RF discharge lamp \cite{zhang2002,daerr2011} and sent into the Xe gas under study.  In our work, however, we use an RF discharge within the primary Xe gas itself. Consequently, the $6s[3/2]_{1}$ auxilliary state in the atoms of interest is excited by the RF discharge, as well as by 147 nm photons emitted from other nearby Xe atoms. 

The $6s[3/2]_{2}$ metastable state is also indirectly excited by the RF discharge, so the total Xe* density includes contributions from this RF excitation as well as optical pumping from the $6s[3/2]_{1}$ auxilliary state. We are able to identify these two independent contributions by comparing 823 nm probe absorption with and without the 916 nm control field. The increase in population of the metastable state with the 916 nm control field applied is what we refer to as ``optically enhanced'' production of Xe*. In our experiment we measured a factor of 11 increase in Xe* density with control field powers of only a few mW. This dramatic enhancement indicates that this general process may be useful in applications where large metastable state densities are desired.

The combination of RF excitation and optical pumping is also used in the relatively new field of optically pumped rare gas lasers (OPRGL's) \cite{han2012}.  There, however, the idea is to optically pump directly {\em from} the metastable state to a higher-lying $p$ state, which then decays to a lower-lying $p$ state via collisional mechanisms. In essence, the goal of OPRGL's is to create a population inversion above the metastable state.  In direct contrast, we optically pump {\em into} the metastable state, with a goal of maximum metastable state population.

\begin{figure}[t]
\includegraphics[width=3.25in]{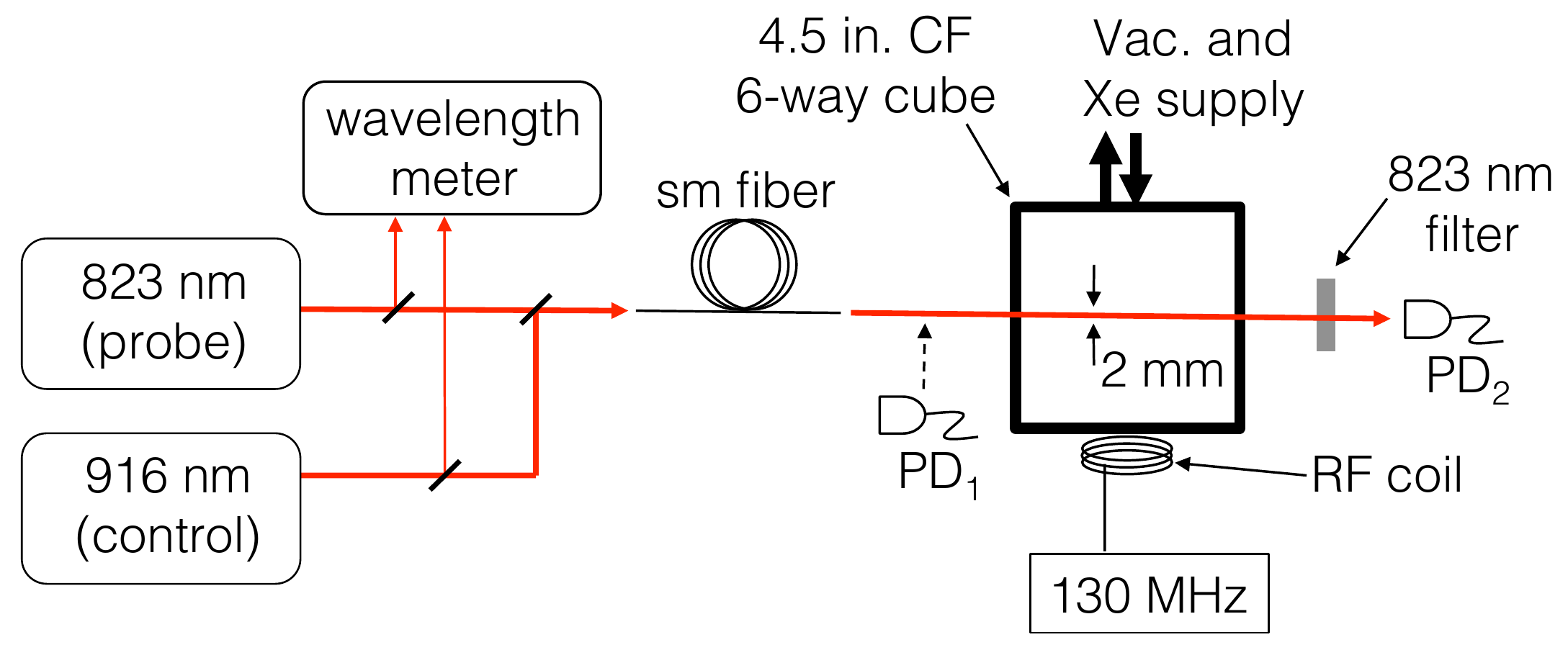}
\vspace*{-.1in}
\caption{(Color online) Overview of the experiment. Photodiodes PD$_{1}$ and PD$_{2}$ were used to measure the power and transmission of the 823 nm probe and 916 nm control beams. An 823 nm narrow bandpass filter was used to isolate the probe from the control.}
\label{fig:fig2}
\end{figure}

Figure \ref{fig:fig2} shows an overview of the experimental set up. Two narrowband continuous-wave tunable diode lasers (Toptica DL pro; $\Delta\nu \sim\!300$ kHz) were used for the 916 nm control and 823 nm probe beams. They were spatially filtered and mode-matched using a common single-mode (sm) fiber, and launched as a free-space beam with a $\sim\!2$ mm diameter through the Xe system. The Xe system consisted of a standard 4.5 inch ConFlat (CF) 6-way cube with optical viewports. The RF discharge was produced using a standard resonant tank circuit consisting of a coil and capacitors (RF frequency = 130 MHz; RF power  $\sim\!25$ W) placed outside of a glass viewport. The cube was typically pumped down to $\sim\!10^{-6}$ torr, and then back-filled with Xe pressures of several torr. The experiment consisted of producing a discharge in the system, and then measuring the 823 nm probe absorption as a function of probe and control detunings ($\Delta_{p}$ and $\Delta_{c}$ in Figure \ref{fig:fig1}).

\begin{figure}[b]
\includegraphics[width=3.25in]{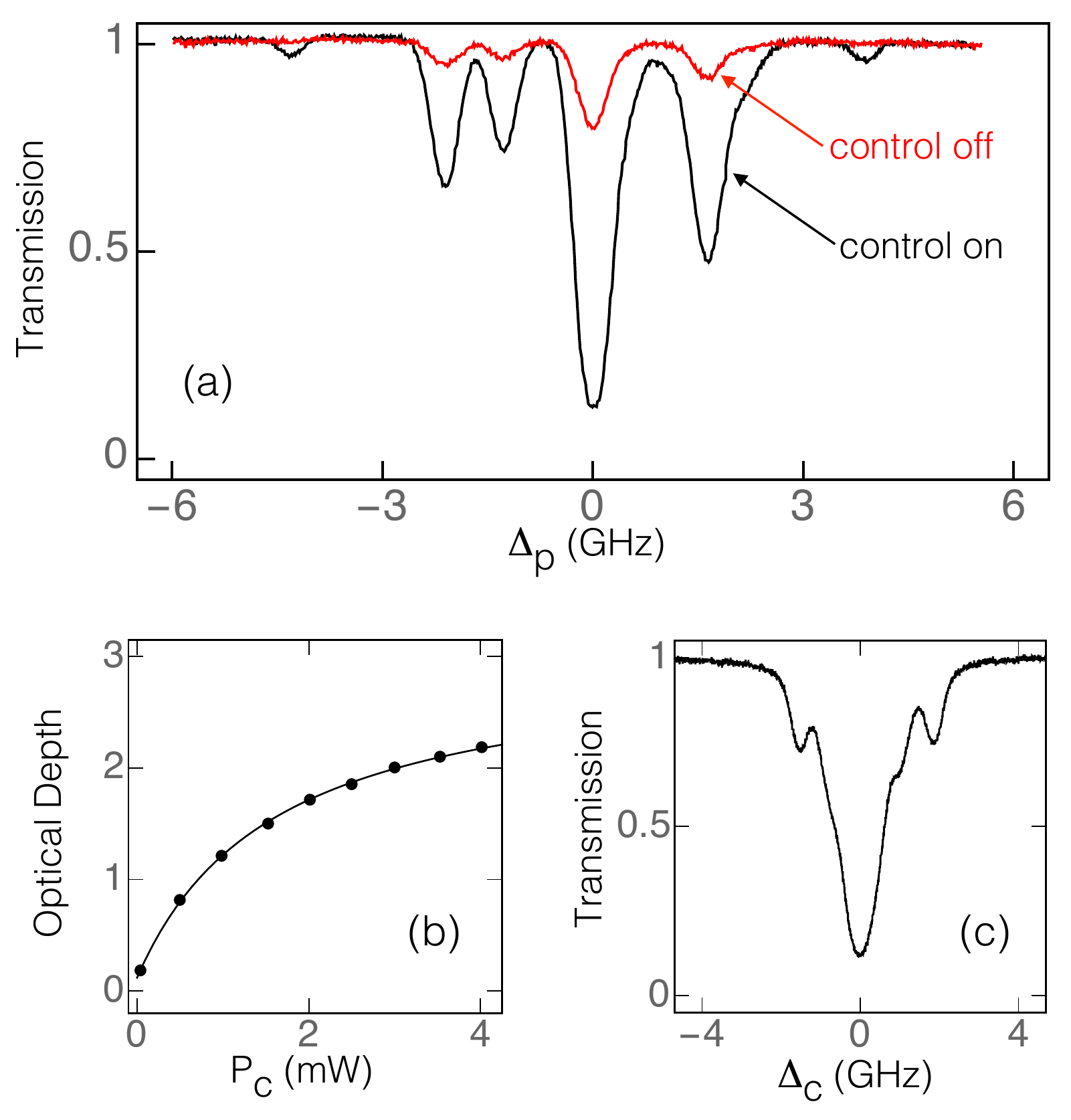}
\vspace*{-.1in}
\caption{(Color online) Demonstration of optically enhanced production of metastable xenon: (a) 823 nm probe transmission spectrum without the 916 nm control  (red curve) and with the 916 nm control  (black curve). (b) 823 nm probe Optical Depth as function of 916 nm control power, $P_{c}$. (c) normalized 823 nm probe transmission as a function of 916 nm control detuning, $\Delta _{c}$.}
\label{fig:fig3}
\end{figure}

Figure \ref{fig:fig3} shows a summary of the experimental results. For all of the data, a weak 823 nm probe beam power of $P_{p} = 7 \: \mu$W was used. First, the red curve in panel (a) shows the 823 nm probe beam transmission spectrum with the RF discharge active but the 916 nm control beam turned off. The 6 absorption dips are produced by the 9 stable isotopes in natural Xe; the large central dip at $\Delta_{p}=0$ is due to a combination of the 7 even isotopes, while the 5 other dips are due to the hyperfine splittings of $^{129}$Xe and $^{131}$Xe \cite{xia2010}. Next, the black curve in panel (a) shows the same 823 nm probe scan with the 916 nm control beam turned on at our maximum available power of $P_{c}=4$ mW.  The significantly increased 823 nm probe absorption is due to the optically enhanced metastable state density, and represents the main result of this paper.

Panel (b) quantifies this optical enhancement by showing the optical depth (OD) experienced by the probe beam (at $\Delta_{p}=0$) as a function of the control field power, $P_{c}$ (with $\Delta_{c}=0$). The data shows an increase from OD = 0.2 at $P_{c}=0$ up to OD = 2.2 at $P_{c}=4$ mW, corresponding to a factor of 11 increase in the Xe* density due to optical enhancement. Here,   91\% of the total metastable state density is due to optical pumping from the  $6s[3/2]_{1}$ auxilliary state, with only 9\% due to standard RF excitation of the metastable state itself.  The domination of the optically-enhanced contribution is largely due to the efficiency of the resonant ``all-optical'' process involving the 147 nm photons \cite{young2002}.

The data in panel (b) also indicates a saturation effect with 916 nm control field power.   The solid curve is a fit to the data by a generic saturation model with a best-fit ultimate OD of 2.9 under these specific conditions.  However, care must be taken not to over-interpret this model, as the apparent saturation is complicated by a number of factors including hyperfine pumping \cite{xia2010}, and the dynamics between optical (147 nm) and RF excitation of the $6s[3/2]_{1}$ auxilliary state.

Panel (c) demonstrates the resonant nature of the optical pumping effect. The data shows the normalized transmission of the 823 nm probe beam (at $\Delta_{p}=0$) as a function of the 916 nm control field detuning (with $P_{c} = 4$ mW).  It can be seen that the optical enhancement in Xe* density rapidly vanishes when the 916 nm control field detuning is larger than the characteristic Doppler-broadened linewidth ($\sim\!750$ MHz) of the $6s[3/2]_{1} \rightarrow 6p[3/2]_{1}$ transition in our system.

The data shown in Figure \ref{fig:fig3} was obtained using a Xe pressure of 6 torr.  We observed qualitatively similar results using Xe pressures in the range of $\sim\! 1 \rightarrow 10$ torr. Lower pressures generally led to smaller relative optical enhancement factors, whereas higher pressures led to smaller overall OD's (i.e. Xe* densities) in our system \cite{xia2010}.  Both of these trends were most likely influenced by the particular geometry of our experiment, where the use of higher Xe pressures tended to confine the RF discharge glow closer to the RF coil, and thus farther from the optical beam path (see Fig. \ref{fig:fig2}).

In summary, long-lived metatstable states of noble gas atoms have spectroscopic properties that are similar to ground state alkali atoms, and there are a number of applications where large metastable densities are desirable \cite{pittman2013,han2012}. Here we have demonstrated the idea of optically enhanced production of metastable xenon by supplementing weak internal RF discharges with strong resonant optical pumping from an auxilliary state. This technique essentially leverages the efficiency of the all-optical production method \cite{young2002}, and may be particularly useful in systems where weak RF discharges can be tolerated and bright external VUV lamps \cite{daerr2011} are not available.

This work was supported by the National Science Foundation under grant No. 1402708.



\begin{thebibliography}{50}

\bibitem{lu2014} Z.-T. Lu, P. Schlosser, W.M. Smethie, N.C. Sturchio, T.P. Fischer, B.M. Kennedy, R. Purtshcert, J.P. Severinghaus, D.K. Solomon, T. Tanhua, and R. Yokochi,   ``Tracer applications of noble gas radionuclides in the geosciences'',  Earth-Sci. Rev. {\bf 138}, 196 (2014).

\bibitem{vassen2012} W. Vassen, C. Cohen-Tannoudji, M. Leduc, D. Boiron, C.I. Westbrook, A. Truscott, K. Baldwin, G. Birkl, P. Cancio, and M. Trippenbach, ``Cold and trapped metastable noble gases'',  Rev. Mod. Phys. {\bf 84}, 175 (2012).

\bibitem{pittman2013} T.B. Pittman, D.E. Jones, and J.D. Franson, ``Ultralow-power nonlinear optics using tapered optical fibers in metastable xenon”, Phys. Rev. A {\bf 88}, 053804 (2013).

\bibitem{hickman2015} G.T. Hickman, T.B. Pittman, and J.D. Franson, “Low-power cross-phase modulation in a metastable xenon-filled cavity for quantum information applications”, Phys. Rev. A {\bf 92}, 053808 (2015).

\bibitem{chen2001} C.Y. Chen, K. Bailey, Y.M. Li, T.P. O’Connor, Z.-T. Lu, X. Du, L. Young and G. Winkler,  ``Beam of metastable krypton atoms extracted from a rf-driven discharge'', Rev. Sci. Instrum. {\bf 72}, 271 (2001).

\bibitem{young2002} L. Young, D. Yang, and R.W. Dunford, ``Optical production of metastable krypton'', J. Phys. B. {\bf 35}, 2985 (2002).

\bibitem{ding2007} Y. Ding, S.-M. Hu, K. Bailey, A.M. Davis, R.W. Dunford, Z.-T. Lu, T. P. O'Connor, and L. Young,  ``Thermal beam of metastable krypton atoms produced by optical excitation'', Rev. Sci. Inst. {\bf 78}, 023103 (2007).

\bibitem{kohler2014}  M. Kohler, H. Daerr, P. Sahling, C. Sieveke, N. Jerschabek, M.B. Kalinowski, C. Becker, and K. Sengstock, ``All-optical production and trapping of metastable noble-gas atoms down to the single-atom regime'', Europhysics Lett. {\bf 108}, 13001 (2014).

\bibitem{NISTASD} A. Kramida, Y. Ralchenko, and J. Reader, NIST Atomic Spectra Database, http://physics.nist.gov/asd (2016).

\bibitem{zhang2002} S. Zhang and S. Zhu, ``Study of xenon discharges for 147 nm emission'', Phys. Scripta {\bf 66}, 476 (2002).

\bibitem{daerr2011} H. Daerr, M. Kohler, P. Sahling, S. Tippenhaure, A. Arabi-Hashemi, C. Becker, K. Sengstock, and M.B. Kalinowski, ``A novel vacuum ultra violet lamp for metastable rare gas experiments'', Rev. Sci. Inst. {\bf 82}, 073106 (2011).

\bibitem{han2012} J. Han and M.C. Heaven, ``Gain and lasing of optically pumped metastable rare gas atoms'', Opt. Lett. {\bf 37}, 2157 (2012).

\bibitem{xia2010} T. Xia, S. W. Morgan, Y.-Y. Jau, and W. Happer, ``Polarization and hyperfine transitions of metastable $^{129}$Xe in discharge cells'', Phys. Rev. A {\bf 81}, 033419 (2010).





\end{thebibliography}
\end{document}